# Abrupt PN Junctions: Analytical Solutions Under Equilibrium and Non-equilibrium

Sina Khorasani

*Abstract*— **We present an explicit solution of carrier and field distributions in abrupt PN junctions under equilibrium. An accurate logarithmic numerical method is implemented and results are compared to the analytical solutions. Analysis of results shows reasonable agreement with numerical solution as well as the depletion layer approximation. We discuss extensions to the asymmetric junctions. Approximate relations for differential capacitance and I-V characteristics are also found under non-zero external bias.**

## I. Introduction

THE semiconductor PN junction forms the basis of all devices being used in the technological developments of microelectronics and optoelectronics industries [1,2]. The theory behind the operation of PN junctions (and extensions to multijunctions such as bipolar transistors) dates back to the 1940s, when William Shockley [3] at the AT&T Bell laboratories devised a basic theoretical framework for understanding of these semiconductor devices [3,4].

Since then, the so-called depletion layer approximation has remained essentially unchanged, despite many shortcomings in correct description of carrier and charge densities as well as electric fields. A number of authors have tried to revisit the basic theoretical formulation of PN junctions to obtain a more accurate description of various parameters [5-7]. This is while, accurate numerical solution of carrier and charge densities still remains a tough procedure to deal with, since electron and hole densities vary over many orders of magnitude across a distance as short as a few microns or less. For a typical junction, these densities may easily vary within up to a factor of $10^{15}$ to $10^{18}$. Hence, satisfaction of boundary and continuity equations at once is not achievable by most of the routine numerical schemes.

A variation of PN junction theory was developed in 2006 [8] without making use of the depletion layer approximation, in which an approximate solution was found using introduction of an auxiliary unphysical parameter. However, no closed form solutions were found and the major contribution was only a modified numerical method to achieve stability and better accuracy.

The recent development of 2D devices, has necessitated access to a better theoretical framework for understanding of submicron-sized PN junctions [9] and bipolar [10,11] devices. In [11], the effects of generation and recombination currents for such 2D devices are fully taken into account. The authors of [12] have also just reported an improved theory for calculation of fields in such devices.

Among the other related papers, we may also mention a few which present analytical discussions of the p-n junctions. Corkisha and Green [13] have calculated the recombination and generation currents in bulk step junctions. Arbitration of doping gradients with carefully engineered profiles has been shown to significantly improve [14] the performance of bipolar transistors. Barbyn and Santos [15] have made an experimental study of p-n junctions under forward bias and through extensive modeling and harmonic analysis within the approximation of depletion layer. A similar harmonic analysis of laser diodes has been conducted by the author [16,17] in the past, which establishes the existence of optimal modulation and mixing in laser diodes. Yang and Schroder [18] have discussed the band bending diagrams and quasi-Fermi levels, and their applications to field-effect transistors and other single- or multi-junction devices.

A recent paper of Boukredimi and Benchouk [19] presents an interesting theoretical study of C-V relationship in symmetric p-n junctions. However, they do not present expressions for the electric and potential field profiles, carrier concentrations, or total charge densities. Moreover, in their analysis the external bias is simply treated as a change in intrinsic carrier density; this assumption appears to be incorrect.

The focus of this paper is to address the basic problems with the theoretical formulations of PN junctions in depth, using nonlinear analysis of governing drift-diffusion kinetics. We start by integration of highly nonlinear master equations for distribution of charges and fields, without strict dependence on the old ideas. Then we succeed to find closed form solutions in terms of relatively simple functions for the case of symmetric step junction under equilibrium. Numerical examples are solved and we show the consistency of our results. We also discuss how to use this method for calculation of parameters in one-sided junctions in which the doping of one side exceeds the other side by a few orders of magnitude.

## II. Theory

### A. Symmetric Junctions

Consider a simple PN junction with uniform doping across the $x = 0$ boundary, as shown in Fig. 1. We assume that the concentration of extrinsic acceptor and donor dopants are

S. Khorasani is with the School of Electrical Engineering, Sharif University of Technology, Tehran, Iran (e-mail: khorasani@sina.sharif.edu).

given by $N_A$ and $N_D$ respectively in the P and N sections. In what follows, we suppose that all dopant and carrier concentrations are normalized to the intrinsic carrier density $n_i$. Hence, for a symmetric junction we may define the dimensionless density $N = N_A/n_i = N_D/n_i$. Similarly, electron and hole carrier densities are normalized accordingly as $n(x) = N(x)/n_i$ and $p(x) = P(x)/n_i$. This allows us to write $n(x)p(x) = 1$ under equilibrium.

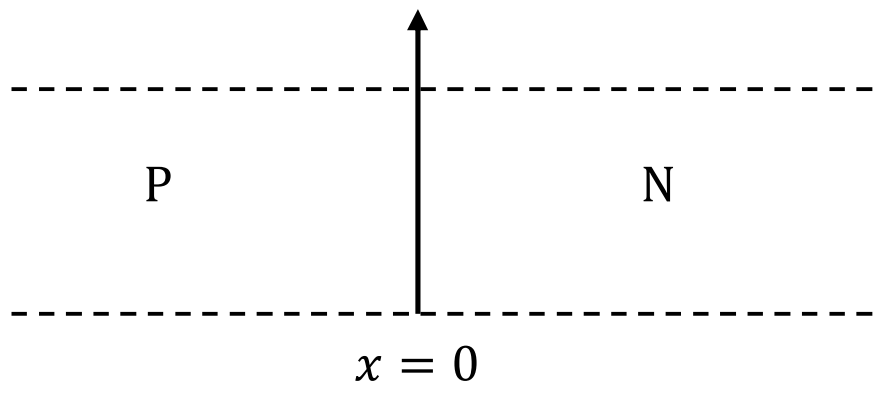

Fig. 1. Illustration of PN junction.

The governing equations for carrier concentrations follow the vanishing sum of drift and diffusion current densities, which result in

$$V_T p'(x) = +\mathcal{E}(x)p(x),$$
$$V_T n'(x) = -\mathcal{E}(x)n(x),$$
(1)

where $V_T = kT/q$ is the thermal voltage with $q$, $T$ and $K$ respectively being the electronic charge, absolute temperature, and Boltzmann's constant. Also, $\mathcal{E}(x)$ is the local electric field, obeying the Poisson equation given by

$$\mathcal{E}'(x) = V_T\delta[p(x) - n(x) - N], \quad x < 0,$$
$$\mathcal{E}'(x) = V_T\delta[p(x) - n(x) + N], \quad x > 0,$$
(2)

in which $\delta = qn_i/\epsilon V_T$ with $\epsilon$ being the permittivity. We define the two conjugate functions

$$\rho(x) = p(x) - n(x),$$
$$\varrho(x) = p(x) + n(x),$$
(3)

which upon insertion in (1) results in

$$\frac{\rho'(x)}{\varrho(x)} = \frac{\varrho'(x)}{\rho(x)} = \frac{\mathcal{E}(x)}{V_T}.$$
(4)

Integration of the above gives

$$\rho^2(x) - \rho^2(-\infty) = \varrho^2(x) - \varrho^2(-\infty).$$
(5)

But we have $\varrho^2(-\infty) - \rho^2(-\infty) = 4n(-\infty)p(-\infty)$ according to (3). Hence, after using $n(x)p(x) = 1$ we obtain

$$\varrho^2(x) = \rho^2(x) + 4.$$
(6)

Now, we first limit ourselves to the case of symmetric junction and solve (1) and (2) only for the $x < 0$ half-space. We initially have

$$\frac{d}{dx}\left[\frac{\rho'(x)}{\varrho(x)}\right] = \delta[\rho(x) - N].$$
(7)

This second-order differential equation is not twice integrable, however, by defining $\alpha = \sqrt{\delta}/\sqrt{N}$ we may obtain the first order approximate equation (for a detailed justification please c.f. Appendix A)

$$\frac{\rho'(x)}{\varrho(x)} = \alpha[\rho(x) - N].$$
(8)

Further substitution of (6) in (8) gives the first-order differential equation

$$\frac{\rho'(x)}{\sqrt{\rho^2(x) + 4}} = \alpha[\rho(x) - N].$$
(9)

This equation can now be integrated exactly. In order to do so, it should be first rearranged as $d\rho/\left[(\rho - N)\sqrt{\rho^2 + 4}\right]$ on the left-hand-side and $\alpha dx$ on the other. This will yield

$$\frac{1}{M}\ln\left[\frac{N - \rho(t)}{M\sqrt{\rho^2(t) + 4} + N\rho(t) + 4}\right]\Bigg|_{t=0}^{t=x} = \alpha t|_{t=0}^{t=x},$$
(10)

with the definition $M = \sqrt{N^2 + 4}$. The correctness of (10) may be verified by straightforward differentiation and some lengthy algebraic simplifications, leading back to (9); this is shown in Appendix B. For the moment, assumption of symmetry requires that $n(0) = p(0)$ and therefore $\rho(0) = 0$. However, the case of strongly asymmetric junctions will be also discussed later. This will result in

$$\frac{N - \rho(x)}{M\sqrt{\rho^2(x) + 4} + N\rho(x)} = \frac{N}{2M}e^{x/L}, \quad x < 0,$$
(11)

where

$$L = \frac{1}{\alpha M},$$
(12)

is a characteristic length of decay (the Debye length is usually denoted by $L_D$ in the literature [1], but we omit the index throughout for the sake of convenience). As we shall see, the behavior of exponential terms in the solution will suggest that it would be actually equivalent to the Debye shielding, or charge screening phenomenon, and hence the effective Debye length will become

$$\ell = \sqrt{\frac{\epsilon V_T}{qn_i N_{\text{eff}}}},$$
(13)

in which we have made use of the approximation $N \cong M$, and the effective density $N_{\text{eff}} = N/\ln\sqrt{N}$ is defined by comparison to numerical simulations (Appendix C).

The equation (11) may be rearranged to a quadratic equation in terms of the function $\rho(x)$, which admits two explicit solutions for $\rho(x)$. In order to do this, the expression should be first re-arranged to have $\sqrt{\rho^2(x) + 4}$ on one side and all other terms on the other side, and then be squared. This would yield a second-order algebraic equation in terms of $\rho(x)$ which can be easily solved. The physically meaningful solution is obtained after some algebra and simplification as

$$\rho(x) = N\frac{1-\frac{4}{N}f(x)-f^2(x)}{1+Nf(x)-f^2(x)}, \quad x<0, \tag{14}$$

in which

$$f(x) = \frac{N}{M+2}e^{x/\ell}. \tag{15}$$

Now, electron and hole densities for $x<0$ may be easily found from (3) and (6) as

$$p(x) = \frac{1}{2}\left[\sqrt{\rho^2(x)+4}+\rho(x)\right],$$
$$n(x) = \frac{1}{2}\left[\sqrt{\rho^2(x)+4}-\rho(x)\right]. \tag{16}$$

Because of symmetry, we may use $p(+x)=n(-x)$ to obtain solutions for $x>0$ as well. The results based on (16) and (14) will guarantee smooth transition of hole and electrons across the boundary at $x=0$. The continuity of derivatives are also supported by symmetry, and easily follow (16), by noting that $\rho(0)=0$. To observe this, it is sufficient to take derivatives of (16) and simplify expressions for $n'(0)$ and $p'(0)$. This will readily give $n'(0)=-p'(0)$ which is the desired result. This continuity of solutions across the origin is illustrated in Fig. 2 below.

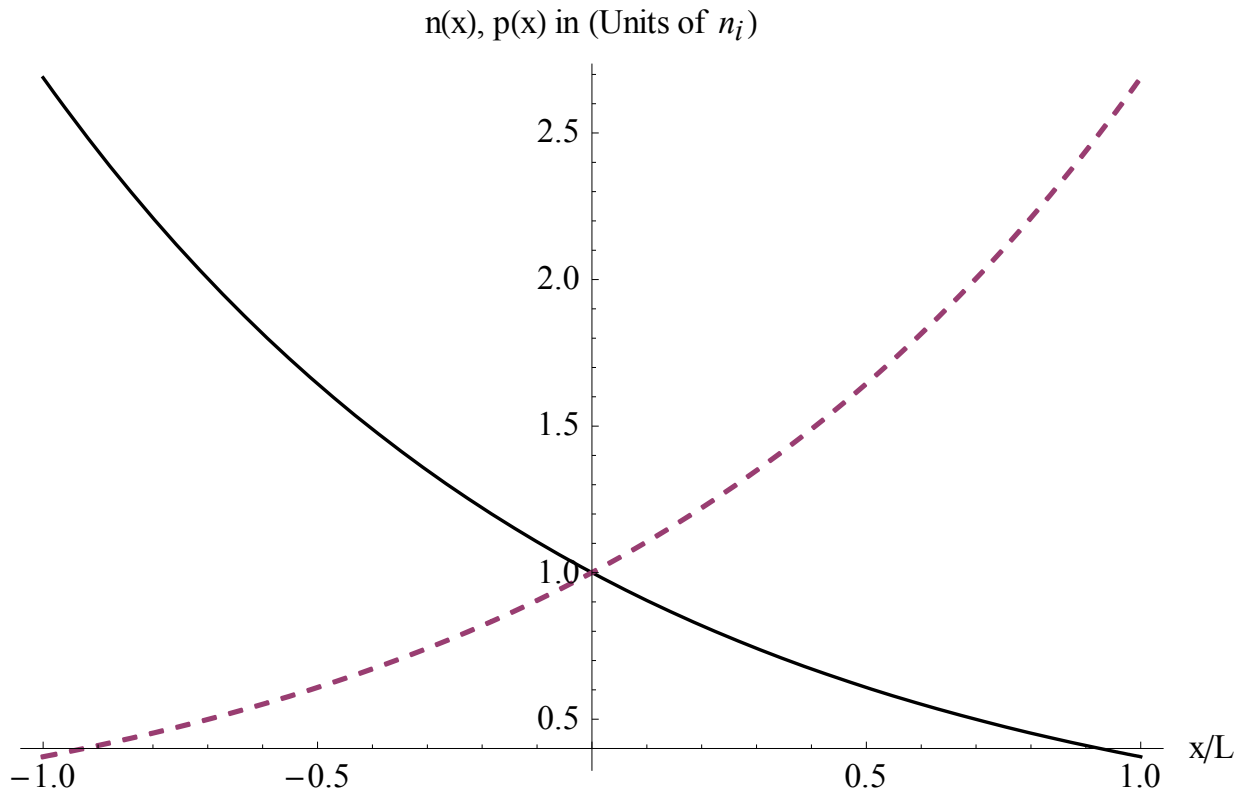

Fig. 2. Smooth transition of carrier densities across the across boundary of the junction: holes (black) and electrons (dashed).

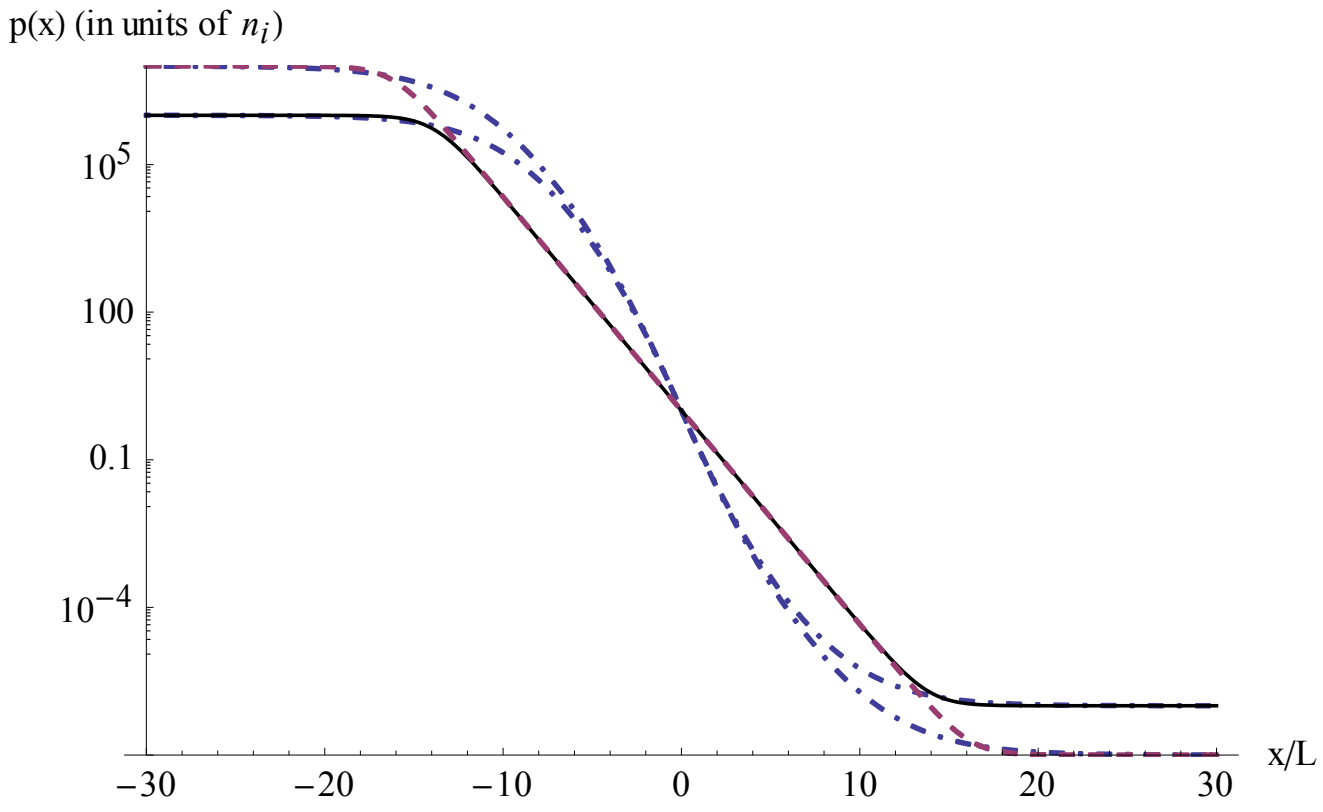

Fig. 3. Concentration of hole carriers for different doping levels: $N=10^6$ or $N_A=N_D=10^6 n_i$, $L=40.1nm$ (black: analytical, dotted: numerical), $N=10^7$ or $N_A=N_D=10^7 n_i$, $L=12.7nm$ (dashed: analytical, dot-dashed: numerical).

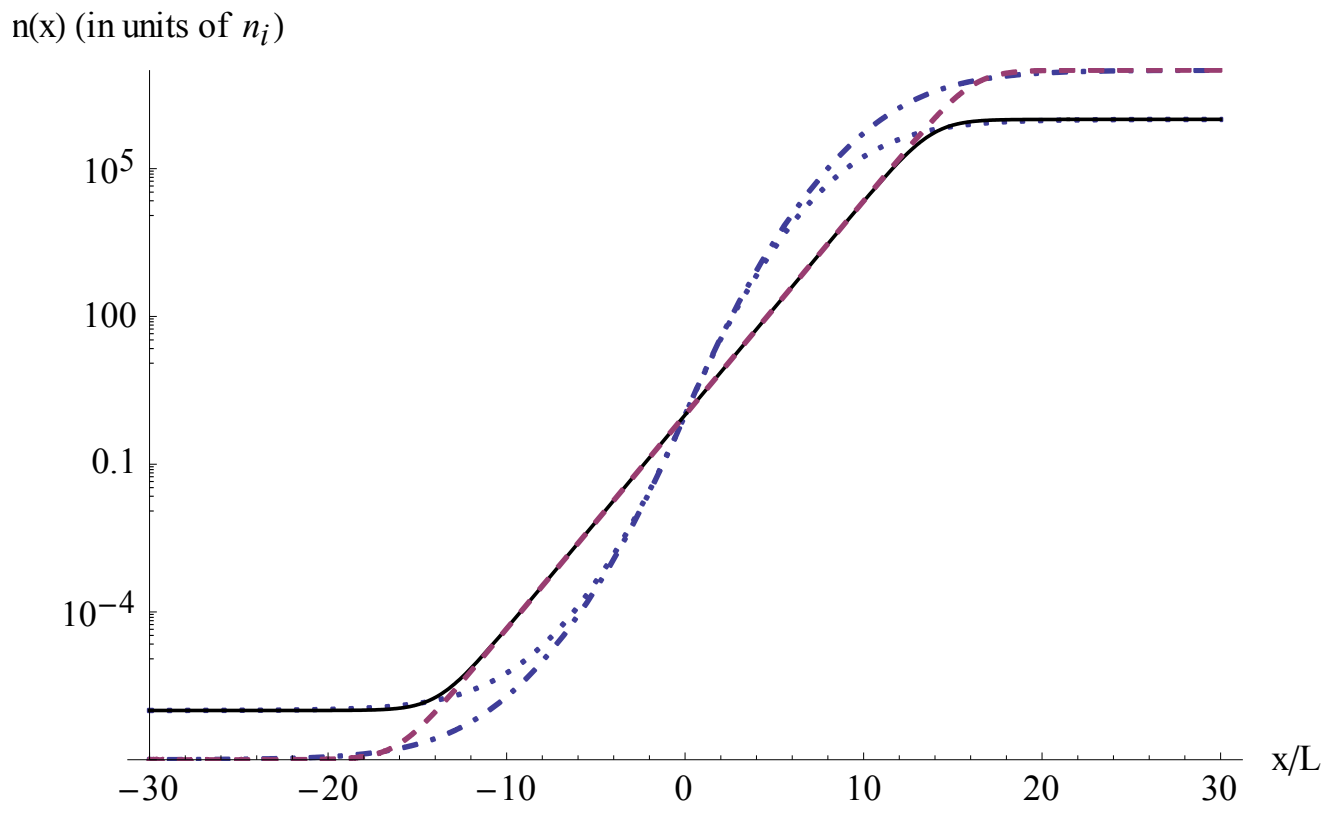

Fig. 4. Concentration of electron carriers for different doping levels: $N=10^6$ or $N_A=N_D=10^6 n_i$, $L=40.1nm$ (black: analytical, dotted: numerical), $N=10^7$ or $N_A=N_D=10^7 n_i$, $L=12.7nm$ (dashed: analytical, dot-dashed: numerical).

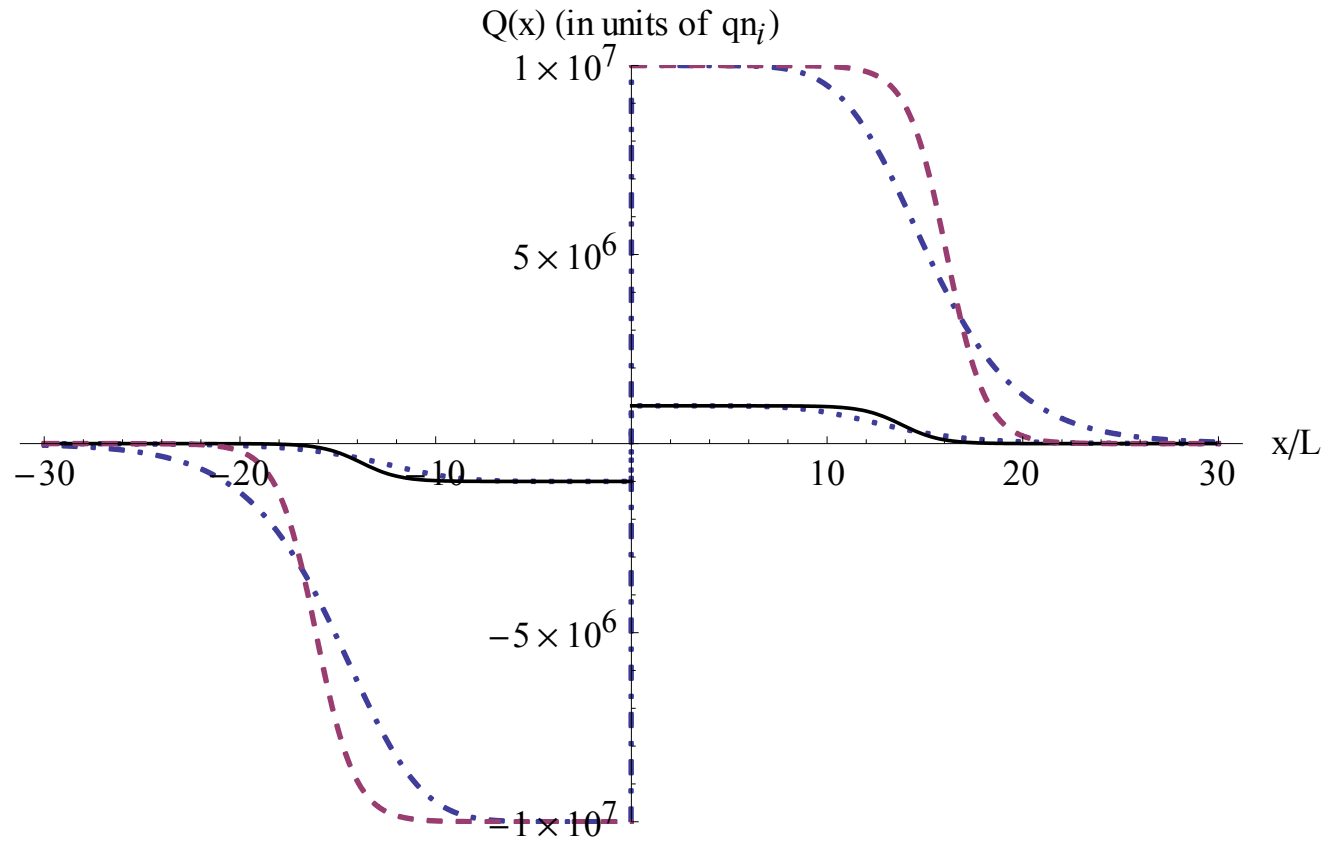

Fig. 5. Total space charge across the junction for different doping levels: $N=10^6$ or $N_A=N_D=10^6 n_i$, $L=40.1nm$ (black: analytical, dotted: numerical), $N=10^7$ or $N_A=N_D=10^7 n_i$, $L=12.7nm$ (dashed: analytical, dot-dashed: numerical).

Both types of carrier densities are plotted for different doping concentrations, too, in Figs. 3 and 4. These plots are shown on the logarithmic scale since the carrier concentrations vary over a multitude of magnitude orders across the junction. It has to be noticed that the Debye length $L$ for each case of the dopant concentrations is different, and the horizontal axis is normalized only for comparison purposes. Similarly, the total space charge across the junction is plotted in Fig. 5 as well. All remains now is to find the electric field $\mathcal{E}(x)$, which may be found by (1); this will however result in relatively large expressions.

So far, everything has been pretty much close to the exact solution. But a very useful alternative form may be found by first noting that $N$ typically ranges between $10^4$ to $10^9$ and is thus much larger than unity. Hence, (14) may be simplified to the very accurate approximation (c.f. Appendix D)

$$\rho(x) = \frac{N}{1-\frac{N}{2}\operatorname{csch}\left(\frac{x}{\ell}\right)}, \quad x<0. \tag{17}$$

Now, one may integrate (17) as

$$\mathcal{E}(x) = V_T\delta \int_{-\infty}^{x} [\rho(t) - N]\, dt, \tag{18}$$

which will lead to the expression for electric field in the half-space $x < 0$

$$\mathcal{E}(x) = -V_T\delta\ell\frac{2N^2}{M}\left\{\tanh^{-1}\left(\frac{N-2}{M}\right) + \tanh^{-1}\left[\frac{2 + N\tanh\left(\frac{x}{2\ell}\right)}{M}\right]\right\}. \tag{19}$$

Using the trigonometric identity $\tanh^{-1}\theta + \tanh^{-1}\varphi = \tanh^{-1}[(\theta + \varphi)/1 + \theta\varphi]$, the above expression may be further simplified as

$$\mathcal{E}(x) = -V_T\delta\ell\frac{2N^2}{M}\tanh^{-1}\left[M\frac{1 + \tanh\left(\frac{x}{2\ell}\right)}{2 + N + (N-2)\tanh\left(\frac{x}{2\ell}\right)}\right]. \tag{20}$$

Evidently, the electric field $\mathcal{E}(x)$ must be an even function so that we have the following full expression, being valid everywhere both for $x < 0$ and $x > 0$

$$\mathcal{E}(x) = -V_T\delta\ell\frac{2N^2}{M}\tanh^{-1}\left[M\frac{1 - \tanh\left|\frac{x}{2\ell}\right|}{2 + N - (N-2)\tanh\left|\frac{x}{2\ell}\right|}\right]. \tag{21}$$

A typical sketch of the electric field is shown in Fig. 6a.

Hence, the maximum electric field $\mathcal{E}_{\max} = |\mathcal{E}(0)|$ at the middle of the junction is given by

$$\mathcal{E}_{\max} = V_T\delta\ell\frac{2N^2}{M}\tanh^{-1}\left[\frac{M}{2+N}\right]. \tag{22}$$

The expression within the brackets is logarithmically divergent and using $\tanh^{-1}(1-z)\sim\frac{1}{2}\ln(2/z)$ for small (and positive) $z$, and taking $1 - z = M/(N+2)$ we obtain $z = 1 - \sqrt{1 - \frac{4N}{(N+2)^2}} \approx 2N/(N+2)^2 \approx 2/N$, where we have used the binomial expansion first, as well as the fact $N \gg 1$ next. After some algebra we get

$$\mathcal{E}_{\max} \cong \frac{V_T}{\ell}\ln N. \tag{23}$$

The built-in voltage of the junction is exactly given by the initial difference in Fermi levels prior to the contact, being

$$V_0 = 2V_T\ln N. \tag{24}$$

Hence, we may define an effective width $W$ of the depletion region as

$$W \equiv \frac{2V_0}{\mathcal{E}_{\max}} = 4\ell. \tag{25}$$

Finally, the voltage or electrical potential $V(x)$across the junction defined as $\mathcal{E}(x) = -V'(x)$ can be obtained by direct integration of (20) which leads to excessively long and inaccurate expression. A much better approximation may be obtained by using the equation (1), which can be readily integrated as

$$V(x) = V_T\ln\left[\frac{N}{p(x)}\right] = V_T\ln\left[\frac{n(x)}{N}\right]. \tag{26}$$

Inserting (17) into (16), and using the above, we obtain an explicit equation for the distribution of electrostatic field. For this purpose, we first extend (17) into both sides of the junction as

$$\rho(x) = -\mathrm{sgn}(x)\frac{N}{1 + \frac{N}{2}\mathrm{csch}\left|\frac{x}{\ell}\right|}, \tag{27}$$

where $\mathrm{sgn}(\cdot)$ is the sign function. A plot of the electric field is shown in Fig. 6b. The tails of the potential variation at the borders of the depletion area are somewhat similar to exponential functions [20].

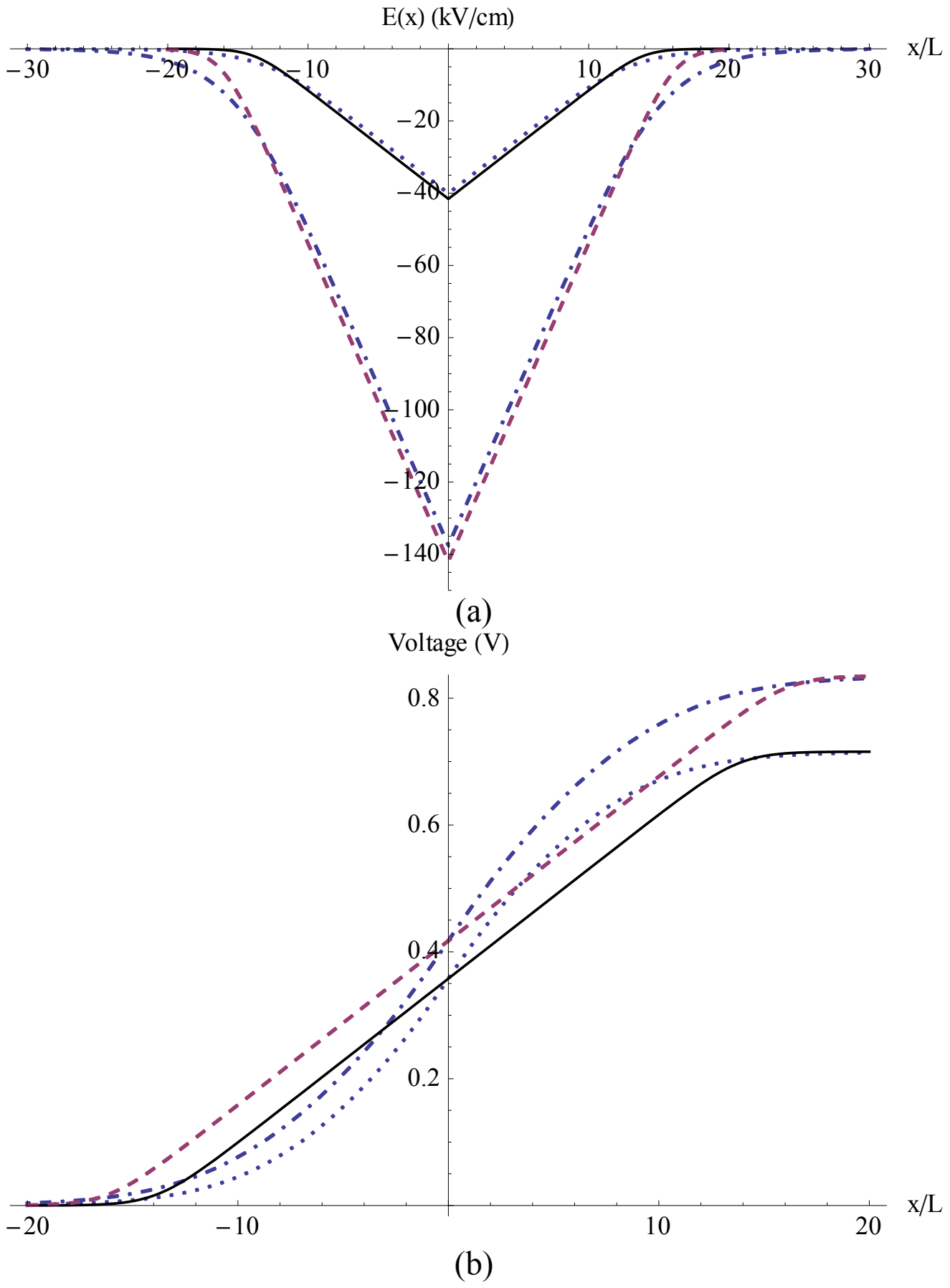

(b)

Fig. 6. Electrostatic field and potential across the junction for difference doping levels: $N = 10^6$ or $N_A = N_D = 10^6 n_i$, $L = 40.1nm$ (black: analytical, dotted: numerical), $N = 10^7$ or $N_A = N_D = 10^7 n_i$, $L = 12.7nm$ (dashed: analytical, dot-dashed: numerical).

### *B. Scaling Laws*

From (23) and (25) we may readily obtain the accurate scaling laws

$$\mathcal{E}_{\max} \propto \sqrt{N \ln N}, \tag{28}$$

for peak electric field and

$$W \propto \sqrt{\frac{1}{N}\ln N}, \tag{29}$$

for effective width of the depletion layer. The results (23) and (25) are indeed in exact agreement with the known expressions from depletion layer analysis.

This result shows that the effective width of the depletion layer actually is inversely proportional to the doping concentration. As a matter of fact, calculations of Fig. 3 also confirm and are in complete agreement with this finding, as by increasing the doping density from $N = 10^6$ to $N = 10^7$, the peak electric field is also increased from 41.6kV/cm to 142kV/cm by a factor of $\sqrt{\frac{7}{6}10}$ which is roughly equal to 3.4.

Figures 7 and 8 compare the exact maximum electric field from Appendix C and effective width of the depletion layer due to our analysis (solid lines) versus depletion layer approximation (dashed lines). Figures 9 and 10 present the relative errors of analytical solution versus the depletion layer approximation given in percent. It is notable that by examining the depletion layer approximation versus our analysis here, it is found that the estimations for the width of depletion layer as well as the maximum electric field coincide fairly well, just within a few percent.

### *C. Highly Asymmetric Junctions*

In this section, we limit ourselves to the case of a highly asymmetric PN+ junction with $N_D \gg N_A$; this is normally the practical case due to existing fabrication procedures in current use. Therefore, we may still define $N = N_A/n_i$ and proceed with (1) as well as

$$\begin{aligned} \mathcal{E}'(x) &= V_T\delta[p(x) - n(x) - N], \quad x < 0, \\ \mathcal{E}'(x) &= 0, \quad x > 0. \end{aligned} \tag{30}$$

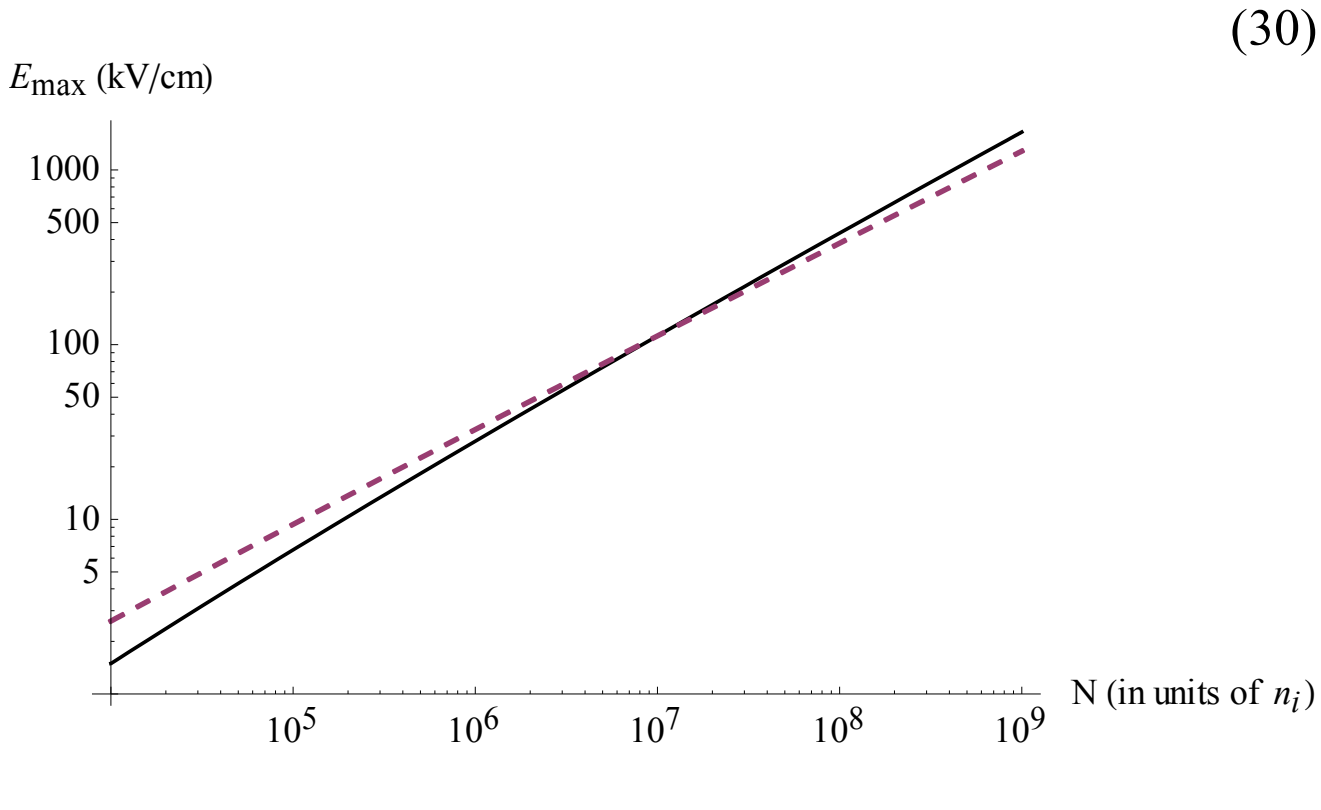

Fig. 7. Dependence of maximum electric field in a symmetric Silicon junction versus dopant concentration: accurate (solid) and simple depletion approximation (dashed).

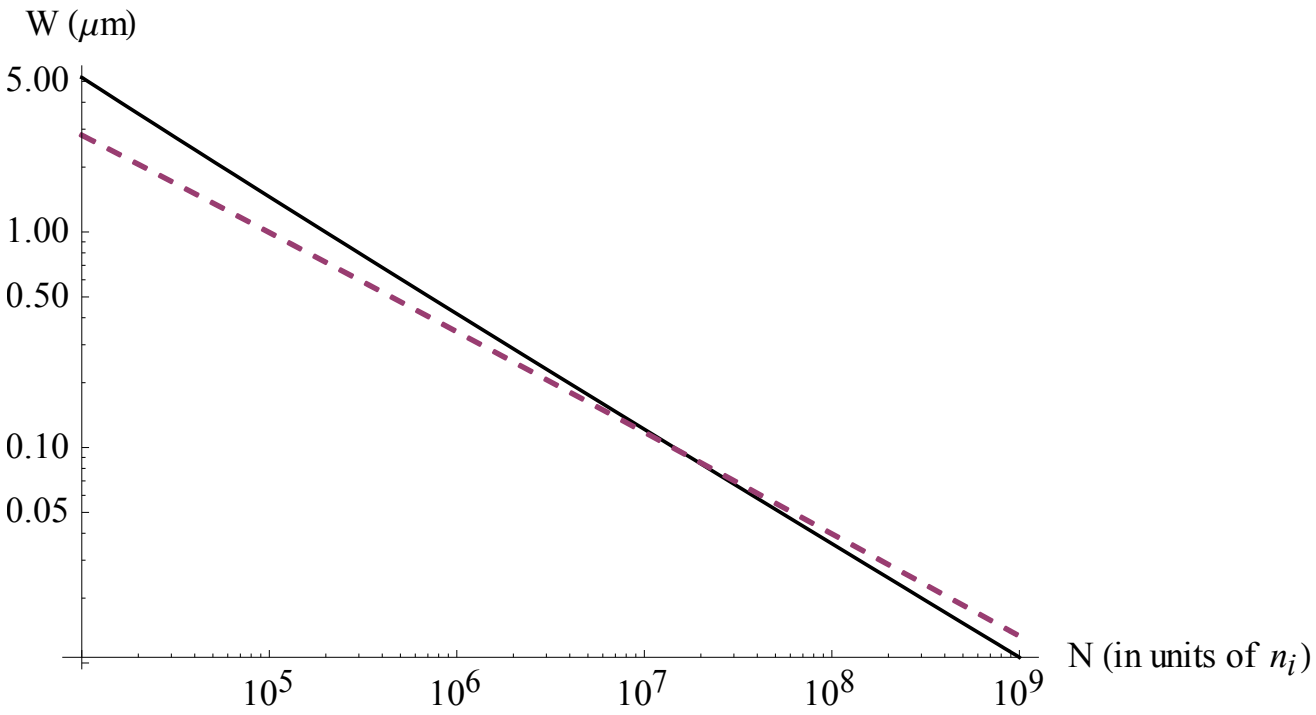

Fig. 8. Dependence of effective depletion width in a symmetric Silicon junction versus dopant concentration: accurate (solid) and simple depletion approximation (dashed).

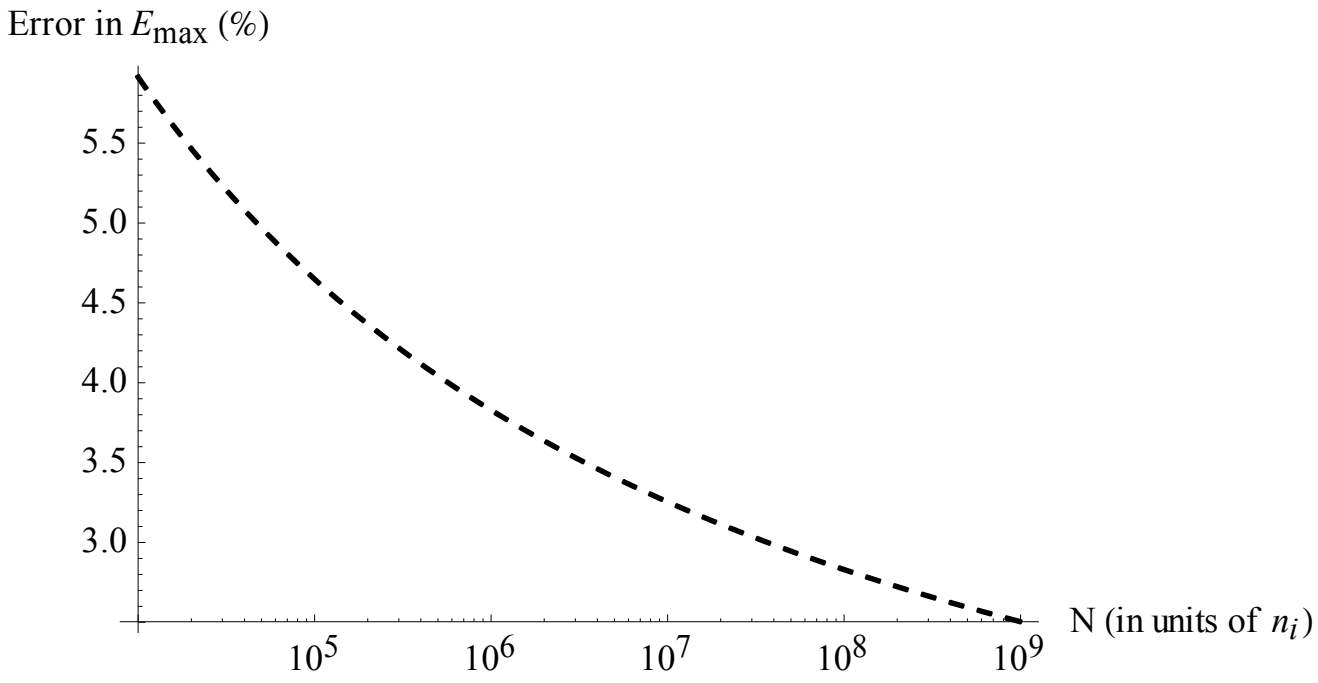

Fig. 9. Relative error in calculation of maximum electric field in depletion layer approximation.

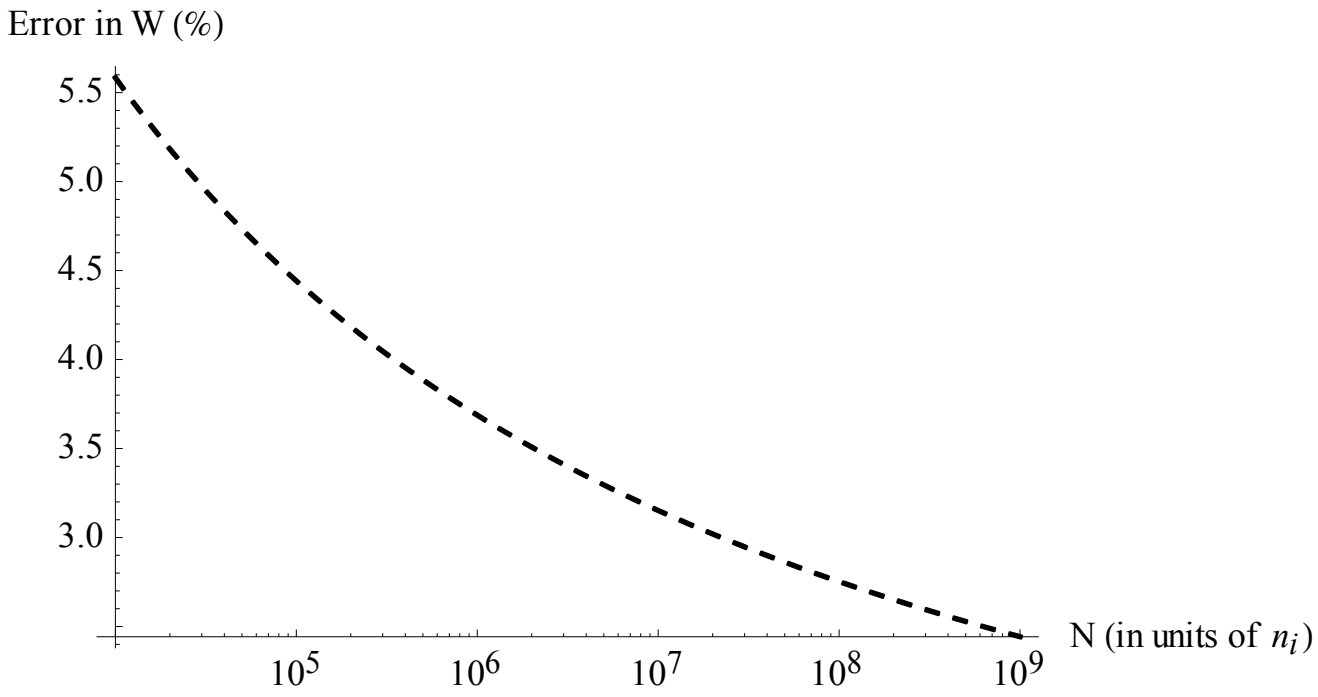

Fig. 10. Relative error in calculation of depletion width in depletion layer approximation.

Justification for (30) follows from the fact that the depletion region is practically pushed out of the highly N+-doped region into the moderately P-doped region. This would imply that during the integration of electric field distribution, which extends just a bit into the highly doped region could be completely ignored. Hence, we have the approximate condition $\mathcal{E}'(x) \approx 0$ in the N+-region because of the almost vanishing contribution of the space-charge. Combined with the boundary condition at infinity $\mathcal{E}(+\infty) = 0$, we then may safely take $\mathcal{E}(x) = 0$ for $x > 0$. This situation is however comparable to a fully symmetric junction with moderate dopings $N_D = N_A$ where the boundary is shifted into the N+-region to the amount of $x_0$ (obtained below), but the depletion layer is abruptly terminated at the origin $x = 0$.

Equation (10) still holds but has to be integrated differently, since $\rho(0)$ is no longer zero. In fact, we may make use of the fact that the whole depletion layer is effectively shifted into the P domain because of the zero curvature of the electric field in the N domain, so that one may have $\rho(0) = \rho_0 < 0$. This will result in

$$\frac{N-\rho(x)}{M\sqrt{\rho^2(x)+4}+N\rho(x)} = \frac{N-\rho_0}{M\sqrt{\rho_0{}^2+4}+N\rho_0}e^{x/\ell}, \quad x<0. \tag{31}$$

By going through a few algebraic steps similar to what was done with (11) to obtain (12), the equation (31) would yield

$$\rho(x) = N\frac{1-\frac{4}{N}g(x)-g^2(x)}{1+Ng(x)-g^2(x)}, \quad x<0,$$

$$g(x) = \frac{2(N-\rho_0)}{M\sqrt{\rho_0{}^2+4}+N\rho_0}e^{x/\ell}. \tag{32}$$

If we define $\rho(x_0)=0$, then $x_0$ can be found from by setting the numerator of $\rho(x_0)$ equal to zero and ignoring the middle term. This will result in

$$\frac{2(N-\rho_0)}{M\sqrt{\rho_0{}^2+4}+N\rho_0}e^{x_0/\ell} \approx \pm 1, \tag{33}$$

out of which the positive sign only may lead to physically meaningful solutions. Hence, we obtain

$$x_0 = -\ell\ln\left[\frac{2(N-\rho_0)}{M\sqrt{\rho_0{}^2+4}+N\rho_0}\right], \tag{34}$$

where by the strong asymmetry we have $\rho_0 = p(0)-n(0)$, and $n(0)\approx N_D/n_i$ and $p(0)\approx n_i/N_D$. Hence, $\rho_0 \approx -N_D/n_i$. Now, by noting $M \approx N = N_A/n_i$ and $N_D \gg N_A$ we get

$$x_0 \approx -\ell\ln\left[\frac{n_i(N_A+N_D)}{N_A{}^2+N_D{}^2}\right] \approx \ell\ln\left[\frac{N_D}{n_i}\right]. \tag{35}$$

Then (32) may be put into the similar accurate form of (17) as

$$\rho(x) = \frac{N}{1-\frac{N}{2}\operatorname{csch}\left(\frac{x-x_0}{\ell}\right)}, \quad x<0. \tag{36}$$

Hence, the expression for the electric field should be modified as

$$\begin{aligned}\mathcal{E}(x) &= -V_T\delta\ell\frac{2N^2}{M}\left\{\tanh^{-1}\left[M\frac{1+\tanh\left(\frac{x-x_0}{2\ell}\right)}{2+N+(N-2)\tanh\left(\frac{x-x_0}{2\ell}\right)}\right]\right.\\ &\left.-\tanh^{-1}\left[M\frac{1+\tanh\left(\frac{x_0}{2\ell}\right)}{2+N+(N-2)\tanh\left(\frac{x_0}{2\ell}\right)}\right]\right\}.\end{aligned} \tag{37}$$

## III. Non-zero Bias

In this section, we study the case of non-equilibrium under external bias. From (D.5), and after numerical tests for large $N$, we may write down the further approximation

$$\rho(x) \approx -2\sinh[\ln N\tanh(x/\mathcal{L})], \tag{38}$$

where $\mathcal{L} = \ell\ln\sqrt{N}$. Hence, we readily obtain by (3) the approximate expressions for equilibrium carrier densities as

$$\begin{aligned}p(x) &\approx \exp[-\ln N\tanh(x/\mathcal{L})],\\ n(x) &\approx \exp[+\ln N\tanh(x/\mathcal{L})].\end{aligned} \tag{39}$$

These expressions consistently satisfy the boundary and initial conditions at origin and infinities. The electric field under equilibrium is given by (1) as

$$\mathcal{E}(x) = V_T\frac{p'(x)}{p(x)} = -\frac{V_T}{\mathcal{L}}\ln N\operatorname{sech}^2(x/\mathcal{L}), \tag{40}$$

resulting in a relation similar to (23) for the maximum electric field. We assume that under positive external bias, (39) are perturbed as

$$\begin{aligned}p(x;V) &\approx \exp\left[-\ln N\tanh\left(\frac{x-\Delta}{\mathcal{L}}\right)\right],\\ n(x;V) &\approx \exp\left[+\ln N\tanh\left(\frac{x+\Delta}{\mathcal{L}}\right)\right],\end{aligned} \tag{41}$$

with $\Delta$ being a voltage-dependent constant to be determined. This will result in

$$\begin{aligned}p(x;V)n(x;V) = \exp\left\{-\ln N\left[\tanh\left(\frac{x-\Delta}{\mathcal{L}}\right)\right.\right.\\ \left.\left.-\tanh\left(\frac{x+\Delta}{\mathcal{L}}\right)\right]\right\}.\end{aligned} \tag{42}$$

Within the depletion region, at $x=0$ we should have

$$p(x;V)n(x;V) = \exp\left(\frac{V}{V_T}\right). \tag{43}$$

This will curiously yield the simple result

$$\Delta = \mathcal{L}\tanh^{-1}\left(\frac{V}{2\ln NV_T}\right) = \mathcal{L}\tanh^{-1}\left(\frac{V}{V_0}\right). \tag{44}$$

We now assume symmetric hole and electron mobilities, and write the equations for the current density $J = 2J_p = 2J_n$ as

$$\begin{aligned}\tfrac{1}{2}J &= q\mu n_i[\mathcal{E}(x;V)p(x;V) - V_Tp'(x;V)],\\ \tfrac{1}{2}J &= q\mu n_i[\mathcal{E}(x;V)n(x;V) + V_Tn'(x;V)].\end{aligned} \tag{45}$$

Combining the above gives

$$\begin{aligned}\mathcal{E}(x;V) &= V_T\frac{p'(x;V)+n'(x;V)}{p(x;V)-n(x;V)},\\ J &= 2q\mu n_i\frac{[p(x;V)n(x;V)]'}{p(x;V)-n(x;V)}.\end{aligned} \tag{46}$$

Using (41) and plugging in (46) we arrive at the non-equilibrium electric field expression given by

$$\mathcal{E}(x;V) = -\frac{V_T}{\mathcal{L}}\ln N \frac{\mathrm{sech}^2\left(\frac{x-\Delta}{\mathcal{L}}\right)p(x;V) - \mathrm{sech}^2\left(\frac{x+\Delta}{\mathcal{L}}\right)n(x;V)}{p(x;V)-n(x;V)}. \tag{47}$$

Hence, by employing hyperbolic-trigonometric identities, and within the quasi-equilibrium approximation, the electric field at $x = 0$ under forward bias is

$$\mathcal{E}(0;V) = -\frac{V_T}{\mathcal{L}}\ln N \,\mathrm{sech}^2\left(\frac{\Delta}{\mathcal{L}}\right) = -\frac{V_T}{\mathcal{L}}\ln N\left[1-\left(\frac{V}{V_0}\right)^2\right]. \tag{48}$$

This expression may be easily generalized for both negative and positive voltages and non-zero current density as

$$\mathcal{E}(0;V) = -\frac{V_T}{\mathcal{L}}\ln N\left[1-\frac{V|V|}{{V_0}^2}\right] + \frac{J}{2q\mu n_i}\exp\left(-\frac{V}{2V_T}\right). \tag{49}$$

*A. Junction Capacitance*

The small-signal capacitance per-unit area of the junction may be found as

$$C = \frac{dQ}{dV}, \tag{50}$$

where $Q$ is the per-unit area charge difference on either side of the junction due to the application of the voltage. It may be evaluated as

$$Q = qn_i \int_{-\infty}^{+\infty} [p(x;V) - p(x;0)]\, dx. \tag{51}$$

Hence, by (50) we get

$$C = qn_i \int_{-\infty}^{+\infty} \frac{\partial}{\partial V} p(x;V)\, dx, \tag{52}$$

which after using (42), (44), and some simplification results in

$$C = \frac{qn_i\mathcal{L}}{V_T\left[1-\frac{V|V|}{{V_0}^2}\right]}\sinh\left(\frac{V_0}{2V_T}\right). \tag{53}$$

Please notice that (53) represents the total capacitive effects of diffusion kinetics and depletion layer combined at once. It is a finite value at zero bias

$$C|_{V=0} = \frac{qn_i\mathcal{L}}{V_T}\sinh\left(\frac{V_0}{2V_T}\right), \tag{54}$$

and diverges at $V = V_0$, approaches zero under strong reverse bias, similar to the well-known behavior of depletion capacitance [1-4].

*B. I-V Characteristics*

Returning to the equation (45) for the current density, if we try to directly plug-in the quasi-equilibrium carrier densities (41), one would obtain zero current. Hence, one would need first to estimate the change in electric field due to bias from another method. For this purpose, we start from the equation

$$\mathcal{E}(y;V) = \mathcal{E}(y;0) + \frac{q}{\epsilon}\int_{-\infty}^{y}[\rho(x;V) - \rho(x;0)]dx, \tag{55}$$

After plugging in (49), this may be re-written as

$$J = \frac{2q^2{n_i}^2\mu}{\epsilon}e^{\frac{V}{2V_T}}\int_{-\infty}^{0}[\rho(x;V) - \rho(x;0)]dx. \tag{56}$$

Evaluation of integral and simplifying results in the expression

$$J(V) = \frac{2q^2{n_i}^2\mu\mathcal{L}}{\epsilon}F(V)e^{\frac{V}{2V_T}}, \tag{57}$$

where the function $F(V)$ is given as

$$F(V) = 2\mathrm{Shi}\left(\frac{V_0}{2V_T}\right)\cosh\left(\frac{V_0}{2V_T}\right) - e^{-\frac{V_0}{2V_T}}\mathrm{Shi}\left[\frac{V_0}{2V_T}\left(1+\frac{V}{V_0}\right)\right] - e^{+\frac{V_0}{2V_T}}\mathrm{Shi}\left[\frac{V_0}{2V_T}\left(1-\frac{V}{V_0}\right)\right]. \tag{59}$$

Here, $\mathrm{Shi}(z)$is defined in terms of the exponential integral as

$$\mathrm{Shi}(z) = \frac{\mathrm{Ei}(z) - \mathrm{Ei}(-z)}{2}, \qquad \mathrm{Ei}(z) = \int_{z}^{\infty}\frac{e^{-u}}{u}du. \tag{60}$$

The dimensionless function $I(V) = F(V)e^{\frac{V}{2V_T}}$ is plotted in Fig. 11.

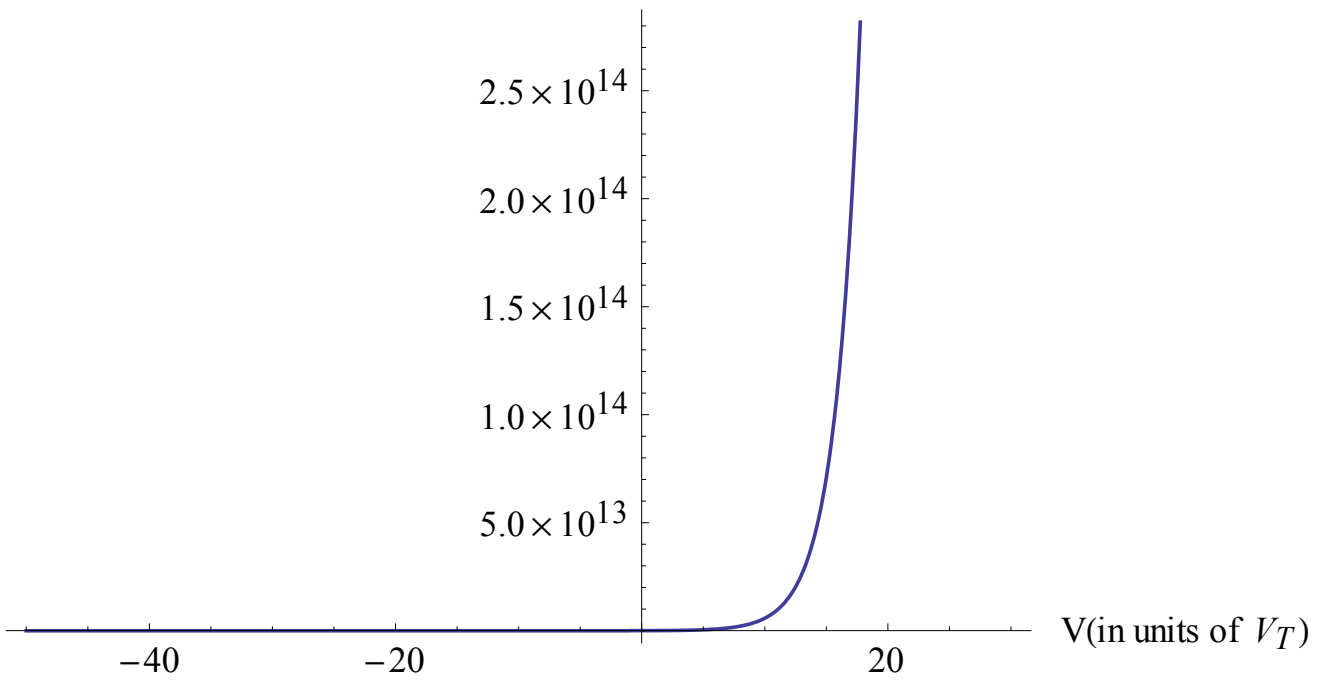

Fig. 11. Normalized approximate I-V characteristics of junction.

## IV. Conclusions

In this paper we discussed a new analytical solution for the carrier and field distributions of an abrupt PN junction under equilibrium. Solutions for symmetric and asymmetric junctions were discussed. We obtained closed form expressions and noticed major differences with respect to the conventional depletion layer approximation. We showed that different accurate scaling laws are expected for the case under study. This will find practical importance in design of compact multi-junction devices. We also obtained approximate expressions for differential capacitance and I-V characteristics.

## Appendix A

The carrier concentrations in the electric field $\mathcal{E}(x)$ equation (2) may be substituted from (1) to obtain

$$\frac{d}{dx}\left[\frac{p'(x)}{p(x)}\right] = +\delta\left[p(x) - \frac{1}{p(x)} - N\right],$$
$$\frac{d}{dx}\left[\frac{n'(x)}{n(x)}\right] = -\delta\left[n(x) - \frac{1}{n(x)} - N\right]. \tag{A.1}$$

These may be integrated once by multiplication of appropriate integration factors $p'(x)/p(x)$ and $n'(x)/n(x)$ to obtain

$$\frac{p'(x)}{p(x)}\frac{d}{dx}\left[\frac{p'(x)}{p(x)}\right] = +\delta\left[p'(x) - \frac{p'(x)}{p^2(x)} - N\frac{p'(x)}{p(x)}\right], \tag{A.2}$$

and similarly for $n(x)$, to find

$$\frac{p'(x)}{p(x)} = -\sqrt{2\delta\left[p(x) + \frac{1}{p(x)} - N\ln p(x) - N - \frac{1}{N} + N\ln N\right]}. \tag{A.3}$$

Here, the integration is taken from $-\infty$ to $x$, while we note that $p(-\infty) = \frac{1}{2}N + \frac{1}{2}\sqrt{N^2+4} \cong N$ and $p'(-\infty) = 0$. The negative root is taken since on the left hand side, the derivative should be negative, while the denominator is positive. This can be reformed as

$$\frac{\rho'(x) + \varrho'(x)}{\rho(x) + \varrho(x)} = -\sqrt{2\delta\left[\varrho(x) - N\ln p(x) - N - \frac{1}{N} + N\ln N\right]}. \tag{A.4}$$

By (4), (6), and (16) we get after some algebra and ignoring the $\frac{1}{N}$ term

$$\frac{\rho'(x)}{\sqrt{\rho^2(x)+4}} = -\sqrt{2\delta N}\sqrt{\frac{\sqrt{\rho^2(x)+4}}{N} - 1 - \ln\left[\frac{\rho(x) + \sqrt{\rho^2(x)+4}}{2N}\right]}. \tag{A.5}$$

Now, we expand the logarithm up to the second order to obtain

$$\ln\left[\frac{\rho(x) + \sqrt{\rho^2(x)+4}}{2N}\right] \sim -\left[1 - \frac{\rho(x) + \sqrt{\rho^2(x)+4}}{2N}\right] - \frac{1}{2}\left[1 - \frac{\rho(x) + \sqrt{\rho^2(x)+4}}{2N}\right]^2 + \cdots. \tag{A.6}$$

By plugging (A.6) in (A.5) and simplifying, we get

$$\frac{\rho'(x)}{\sqrt{\rho^2(x)+4}} = -\sqrt{2\delta N}\sqrt{\frac{\sqrt{\rho^2(x)+4} - \rho(x)}{2N} + \frac{1}{2}\left[1 - \frac{\rho(x) + \sqrt{\rho^2(x)+4}}{2N}\right]^2}. \tag{A.7}$$

Using the approximation $\sqrt{\rho^2(x)+4} \approx \rho(x)$ only on the right hand side, and noting the correct choice of signs in taking the root by $\rho(x) < N$, we finally arrive at

$$\frac{\rho'(x)}{\sqrt{\rho^2(x)+4}} = \sqrt{\frac{\delta}{N}}[\rho(x) - N], \tag{A.8}$$

which is basically having the same form as (9), if we take $\alpha = \sqrt{\delta/N}$.

## Appendix B

Differentiation of (10) yields

$$\frac{d}{d\rho}\left\{\frac{1}{M}\ln\left[\frac{N-\rho}{M\sqrt{\rho^2+4} + N\rho + 4}\right]\right\}$$
$$= \frac{1}{M}\frac{1}{\rho - N} - \frac{1}{M}\frac{N + M\frac{\rho}{\sqrt{\rho^2+4}}}{4 + N\rho + M\sqrt{\rho^2+4}}$$
$$= \frac{1}{M}\frac{1}{\rho - N} - \frac{1}{M}\frac{N\sqrt{\rho^2+4} + M\rho}{\sqrt{\rho^2+4}\left[4 + N\rho + M\sqrt{\rho^2+4}\right]}$$
$$= \frac{1}{M}\frac{1}{\rho - N}\left\{1 - \frac{N\sqrt{\rho^2+4}(\rho - N) + M\rho(\rho - N)}{\sqrt{\rho^2+4}\left[4 + N\rho + M\sqrt{\rho^2+4}\right]}\right\}$$
$$\equiv \frac{1}{M}\frac{1}{(\rho - N)\sqrt{\rho^2+4}}\left\{\frac{R}{S}\right\}. \tag{B.1}$$

The denominator $S$ is given by

$$S = 4 + N\rho + M\sqrt{\rho^2+4}. \tag{B.2}$$

We may expand the numerator $R$ within the braces as

$$R = \sqrt{\rho^2+4}\left[4 + N\rho + M\sqrt{\rho^2+4}\right] - N\sqrt{\rho^2+4}(\rho - N) - M\rho(\rho - N). \tag{B.3}$$

This can be simplified as

$$R = \sqrt{\rho^2+4}[4+N\rho-N(\rho-N)]+M\rho^2+4M-M\rho^2 \\ +MN\rho=\sqrt{\rho^2+4}[4+N^2]+4M+MN\rho \\ =M^2\sqrt{\rho^2+4}+4M+MN\rho \\ =M\left[M\sqrt{\rho^2+4}+4+N\rho\right]. \tag{B.4}$$

Hence, we have $R=MS$, and thus (9) is the true derivative of (10) as asserted above.

## APPENDIX C

The equations (1) and (2) for $x<0$ may be combined to obtain

$$\wp''(x)=\delta[2\sinh\wp(x)-N], \tag{C.1}$$

where $\wp(x)=\ln p(x)$ or $\mathcal{E}(x)=V_T\wp'(x)$. Solution of this nonlinear equation usually leads to numerically unstable results. But it can be analytically integrated once on $(-\infty,x)$ by multiplying both sides by $\wp'(x)$, and noting that $\wp'(x)<0$, $\wp'(-\infty)=0$, and $\wp(-\infty)=\ln N$. This will give the firs-order logarithmic differential equation

$$\wp'(x)=\sqrt{2\delta}\sqrt{2\cosh\wp(x)-2\cosh(\ln N)-N[\wp(x)-\ln N]}, \tag{C.2}$$

which can now be numerically integrated stably subject to the boundary condition $\wp(0)=0$ for $x<0$. Hence, the maximum electric field is exactly given by

$$\mathcal{E}_{\max}=V_T\sqrt{2\delta}\sqrt{2-2\cosh(\ln N)+N\ln N}. \tag{C.3}$$

## APPENDIX D

Referring to the equations (14) and (15) we have for $x<0$

$$\rho(x)=N\frac{1-\frac{4}{N}\left(\frac{N}{M+2}e^{x/\ell}\right)-\left(\frac{N}{M+2}e^{x/\ell}\right)^2}{1+N\left(\frac{N}{M+2}e^{x/\ell}\right)-\left(\frac{N}{M+2}e^{x/\ell}\right)^2}. \tag{D.1}$$

Dividing the numerator and denominator by $e^{x/\ell}$ we find

$$\rho(x)=N\frac{e^{-x/\ell}-\frac{4}{M+2}-\left(\frac{N}{M+2}\right)^2e^{x/\ell}}{e^{-x/\ell}+\left(\frac{N^2}{M+2}\right)-\left(\frac{N}{M+2}\right)^2e^{x/\ell}}. \tag{D.2}$$

Now, we make use of the approximations $M+2\approx N$ for large $N$, and find

$$\rho(x)=N\frac{e^{-x/\ell}-\frac{4}{N}-e^{x/\ell}}{e^{-x/\ell}+N-e^{x/\ell}}, \tag{D.3}$$

which by can be simplified using hyperbolic-trigonometric identities as

$$\rho(x)=N\frac{-2\sinh(x/\ell)-\frac{4}{N}}{-2\sinh(x/\ell)+N}, \tag{D.4}$$

and subsequently as

$$\rho(x)=N\frac{1-\frac{4}{N}}{1+\frac{N}{2}\operatorname{csch}(x/\ell)}. \tag{D.5}$$

All remains is to ignore $4/N$ for large $N$ in the numerator to obtain (17).